\documentclass[sigconf]{acmart}
\usepackage{booktabs} 

\usepackage{mdwlist}  
\usepackage[bottom,multiple]{footmisc} 
\usepackage{balance} 

\copyrightyear{2018}
\acmYear{2018} 
\setcopyright{iw3c2w3}
\acmConference[WWW 2018]{The 2018 Web Conference}{April 23--27, 2018}{Lyon, France}
\acmBooktitle{WWW 2018: The 2018 Web Conference, April 23--27, 2018, Lyon, France}
\acmPrice{}
\acmDOI{https://doi.org/10.1145/3178876.3186094}
\acmISBN{978-1-4503-5639-8/18/04}

\begin{document}
\title{Platform Criminalism}
\subtitle{The `Last-Mile' Geography of the Darknet Market Supply Chain}

\author{Martin Dittus}
\orcid{0000-0001-7810-1520}
\affiliation{%
  \department{Oxford Internet Institute}
  \institution{University of Oxford}
  \city{Oxford}
  \country{UK}
}
\email{martin.dittus@oii.ox.ac.uk}

\author{Joss Wright}
\orcid{0000-0001-5237-3309}
\affiliation{%
  \department{Oxford Internet Institute}
  \institution{University of Oxford}
  \city{Oxford}
  \country{UK}
}
\affiliation{%
   \institution{The Alan Turing Institute}
   \city{London}
   \country{UK}
}
\email{joss.wright@oii.ox.ac.uk}

\author{Mark Graham}
\orcid{0000-0001-8370-9848}
\affiliation{%
  \department{Oxford Internet Institute}
  \institution{University of Oxford}
  \city{Oxford}
  \country{UK}
}
\affiliation{%
   \institution{The Alan Turing Institute}
   \city{London}
   \country{UK}
}
\email{mark.graham@oii.ox.ac.uk}

\renewcommand{\shortauthors}{M. Dittus et al.}

\begin{abstract}
Does recent growth of darknet markets signify a slow reorganisation of the illicit drug trade? Where are darknet markets situated in the global drug supply chain? In principle, these platforms allow producers to sell directly to end users, bypassing traditional trafficking routes. And yet, there is evidence that many offerings originate from a small number of highly active consumer countries, rather than from countries that are primarily known for drug production. In a large-scale empirical study, we determine the darknet trading geography of three plant-based drugs across four of the largest darknet markets, and compare it to the global footprint of production and consumption for these drugs. We present strong evidence that cannabis and cocaine vendors are primarily located in a small number of consumer countries, rather than producer countries, suggesting that darknet trading happens at the `last mile', possibly leaving old trafficking routes intact. A model to explain trading volumes of opiates is inconclusive. We cannot find evidence for significant production-side offerings across any of the drug types or marketplaces. Our evidence further suggests that the geography of darknet market trades is primarily driven by existing consumer demand, rather than new demand fostered by individual markets.
\end{abstract}

%
%



\maketitle

\section{Introduction} 
\label{sec:introduction}
The global illicit drug trade is estimated to be worth USD \$321Bn~\cite{unodc2005world}. Traditionally, drugs have travelled along long and complicated supply chains~\cite{gilbert2017silicon}. However, the emergence of new trading platforms has the potential to change these relationships. Darknet markets, also called cryptomarkets, rely on anonymising technology and electronic currencies to facilitate licit and illicit transactions among participants in relative anonymity~\cite{christin2013traveling,van2013surfing,barratt2016everything,kethineni2017use}. Recent studies suggest that they operate at a comparatively small scale relative to the global drug economy, however there are numerous indications of rapid growth~\cite{aldridge2014not,soska2015measuring,baravalle2016mining,van2016characterising,ladegaard2017we}. 
Similar to their legal equivalents, these illicit markets allow the simultaneous presence of multiple types of actors on the same platform~\cite{martin2014drugs,gilbert2017silicon}. They act as brokers in a multi-sided network between large numbers of supply- and demand-side participants, providing a shared interface for varied interactions that formerly took place on open markets~\cite{langley2016platform}. In principle this introduces opportunities to replace old intermediaries with new entrants, potentially reconfiguring the physical supply chains underlying these markets in the process~\cite{foster2017reconsidering}. While such effects have been documented for commercial platforms~\cite{langley2016platform,foster2017reconsidering}, relatively little is known about their impact on the illicit trade.

In response to such shifts, there is great interest among researchers and policy makers to better understand the emergent phenomenon of darknet markets, including some central aspects of their economic geography. However, current understanding of these platforms is still limited. In particular, there is little published knowledge about their link to public health and public policy. What is the relationship between darknet trades and the global distribution and consumption of illicit drugs? More specifically, what is their place in the global drug supply chain? Answers to these questions help reveal the ways in which a layer of the internet specifically designed to conceal geography might in turn alter material economic geographies.
A secondary research interest relates to the use of darknet market data as as an indicator measure in predictive models of broader societal phenomena. In principle, their trading activity provides an evidence base for drug distribution and consumption behaviours that would otherwise be hard to study at large scale. However, before such use it is first necessary to improve our understanding of the nature and limitations of this evidence, and of its fitness for purpose. 
Is there spatial bias inherent in the data? Does this data allow us to draw inferences about the broader drug supply chain, or only certain specific aspects?

Principles from economic geography allow us to study these aspects in a rigorous and empirical manner. Most importantly, the framework of global production networks (GPN) can be used to understand how global supply chains are formed.\footnote[1]{While we are primarily drawing on the GPN literature, we use the terms \emph{global production network} and \emph{supply chain} interchangeably in this paper.} Central to the theory is the observation that such relationships are inherently spatial: there is a geography of production, distribution, and consumption~\cite{hess2006whither}. For certain goods, it is possible to identify distinct spatial profiles for different parts of the supply chain. In the context of illicit drugs, such spatial profiling is possible for plant-based drugs such as cannabis, cocaine, and opiates~\cite{unodc2017world}. Their site of production can be determined with some precision based on biomarkers, as well as arrest records after large seizures, which makes it possible to distinguish supply and demand geographies in some detail. It is harder to establish similar spatial profiles for synthetic drugs.

Based on these foundations, we present the first study to address the economic geography of darknet markets in a comprehensive empirical manner, with a focus on the supply chain underlying these markets. Our study relates regional darknet trading activity with the supply- and demand-side geographies of the three widely traded plant-based drug types cannabis, cocaine, and opiates. 
Our study seeks to address two primary questions:
\begin{enumerate}
  \item Is the geography of darknet trading sited closer to the supply- or demand-sides of illegal drug production networks? In other words, are darknet markets directly connecting producers to consumers, or are they simply new kinds of intermediaries, leaving existing supply chains in place?
\item Do these relationships vary across markets? In other words, do some darknet markets specialise on particular segments of the supply chain?
\end{enumerate}

We find that darknet markets are strongly characterised by a `last-mile' geography that is focused on a small number of highly active consumer countries, rather than presenting as global platforms that disintermediate existing global production networks. This finding is highly consistent across market platforms. Study outcomes further suggest that darknet trading is largely shaped by global demand, rather than fostering new demand. As a result, it is likely that the economic incentives offered by this external demand yield a high ecological robustness against interventions.

This study makes two further contributions to the literature. A first contribution is conceptual: we introduce concepts from economic geography to frame our inquiry, situate darknet markets in existing platform discourse, and draw on criminology to characterise the nature of illicit supply chains. This grounding of our analysis in existing theory allows us to locate darknet markets in the global illicit supply chain. In a second empirical contribution, we systematically collect a large volume of evidence across four large markets, comprising of 1.5M trades, which represents 80\% of the global inventory at the time of study. Our findings are consistent with a large number of prior studies, and our evidence base allows us to develop further understanding. Based on these factors, we propose that our observations have high external validity. 

\section{Related Work} 
\label{sec:related_work}
According to political economist Nick Srnicek, platforms are ``a newly predominant type of business model premised upon bringing different groups together''~\cite{srnicek2017challenges}. Typical examples of platform businesses include Amazon, Google, Uber, and others. These businesses act as brokers in a multi-sided network between supply- and demand-side participants, providing a shared interface for interactions that formerly took place on open markets. Successful platforms derive competitive advantage from network effects, aggregating large numbers of participants~\cite{langley2016platform}. Their centralising power can allow the largest platforms to achieve a quasi-monopoly position~\cite{srnicek2016platform}. Their business models typically involve rent-seeking in the form of transaction fees, and other forms of profit derived from the hosted transactions and resulting data trails~\cite{langley2016platform}.

The economic power of this arrangement is not constrained to purely legitimate forms of trade. Past research has already documented the platform and service infrastructure of the global spam value chain~\cite{levchenko2011click}. More recently, darknet markets have emerged as a new kind of platform for trades of illicit goods and services. A growing body of research is documenting their activities, however there is still little understanding of their platformed nature. Where are they situated in the global illicit supply chain? 
To address this question, we will first review existing characterisations of darknet markets. We then review characterisations of the illicit supply chains supporting the online drug trade. Finally, we relate these two aspects, with a specific focus on the economic geography of illicit trades. We conclude our review with a summary of open issues.

\subsection{Characterisations of Darknet Markets} 
\label{sub:characterisations_of_darknet_markets}
Silk Road, the first prominent darknet market, introduced many key practices that are still used by today's darknet markets. Known as an `eBay for drugs' and modelled on e-commerce platforms, it relied on anonymising technology and electronic currencies to facilitate transactions in relative anonymity~\cite{christin2013traveling,van2013surfing,barratt2016everything,kethineni2017use}. Ideologically, it was shaped by market-libertarian underpinnings: the stated aim of its founder was to create an open marketplace for any conceivable form of trade, seemingly out of reach of the state~\cite{zajacz2017silk}.

Law enforcement has since managed to shut down Silk Road, but dozens of successor platforms have achieved similar prominence since then, and the overall ecosystem appears to be resilient to law enforcement take-downs: new platforms appear as old ones close, driven by steady customer demand~\cite{soska2015measuring,van2017recovery}. Markets offer an ever-increasing repertoire of illicit goods and services, with a prominent focus on recreational drugs such as cocaine, cannabis, and psychedelics, but also opiates such as heroin, various forms of novel psychoactive substances, and digital information goods such as stolen credit cards and other accounts~\cite{soska2015measuring,baravalle2016mining,van2016sells,dolliver2016geographic,broseus2017geographical}. 

Almost paradoxically, darknet markets are often relatively easily accessed by anyone with a Tor browser, informed by an aspiration to increase adoption~\cite{zajacz2017silk}. Their high volatility makes it difficult to reliably track the overall ecosystem over time, however there are numerous indications of significant growth~\cite{aldridge2014not,soska2015measuring,baravalle2016mining,van2016characterising,ladegaard2017we}. Yet compared to the overall drug trade, trading volumes are still negligible. According to a 2015 estimate, the total turnover across darknet markets was in the order of USD \$150-180M/year~\cite{soska2015measuring}. By comparison, a 2005 report by the UN Office on Drug and Crime (UNODC) estimates the retail value of the global illicit drug trade at USD \$321Bn~\cite{unodc2005world}.

Motivations to participate in darknet trading are varied. Typical participants are young and male~\cite{van2013surfing,barratt2014use,barratt2016safer,barratt2016if,van2016characterising}. While many first-time users are simply driven by curiosity or technical interest~\cite{van2013surfing}, a large share of participants has the intention to engage in regular trade, sometimes with the intention to resell~\cite{van2014responsible,barratt2016if,barratt2016everything}. 
Buyers report `window-shopping' behaviour, where they join a market with a particular purchasing decision, but then begin ordering other goods that are also on offer~\cite{barratt2016if}, illustrating the platformed nature. 
Similar to offline crime, participation in darknet markets incurs various forms of risk, including detection by law enforcement, loss of shipments, vendor fraud, and others~\cite{moeller2017flow}. The perceived benefits include a reduction of physical danger compared to street trades, increased product quality at cheaper prices, speedy delivery, and the convenience of varied and well-presented offerings from the same source~\cite{van2013surfing,van2014responsible,barratt2016safer,barratt2016if}. 

\subsection{Illicit Supply Chains} 
\label{sub:illicit_supply_chains}
Conventional drug distribution networks often rely on hierarchical management structures to coordinate their trade~\cite{hughes2017social}. At the same time, recent research in crime science has portrayed such organised drug trafficking as mostly the work of small groups of loosely linked entrepreneurs, rather than large and deeply structured criminal syndicates~\cite{natarajan2006understanding,wood2017structure}. In principle, such loose arrangements are amenable to large-scale reorganisation. Novel technologies such as anonymising software, cryptocurrencies, and other innovations can act as potential enablers for such a reorganisation~\cite{aldridge2014not,barratt2016everything,gilbert2017silicon}. Do they already have an apparent effect on the structure of the global drug production network? 

There are indicators that darknet market distribution networks are starting to differ from the classic cartel distribution channels of the 20th century, where there was a linear flow from growers, wholesale aggregators and processors, international transporters, intermediate distributors, and retail or hand-to-hand sellers~\cite{gilbert2017silicon}. Instead, darknet markets allow the simultaneous presence of multiple types of actors on the same platforms~\cite{martin2014drugs}. In a review of opioid supply chains, three case studies illustrate how products can be bought, repackaged, and resold on the same platform~\cite{gilbert2017silicon}.

Yet, relatively little is known about the actual distribution networks behind darknet market offerings. Early studies appear to confirm the varied and dispersed nature of organisational models. A study of an online opioid market finds that participant interactions are structured as a small-world network, rather than the more linear model of a traditional supply chain~\cite{duxbury2017building}. These networked relations were found to be shaped by preferential attachment, and a need to develop trusted relationships among anonymous participants. Such structures allow for a diffusion of commerce that is more amendable to new entrants, including bulk traders, potentially supporting subsequent redistribution elsewhere~\cite{aldridge2016hidden}. 

\subsection{Economic Geography of Darknet Markets} 
\label{sub:the_geography_of_darknet_markets}
The supply chains behind darknet trades can be described in terms of their economic geography. To manage risk during shipping, darknet market vendors can state their own location and acceptable delivery destinations in product listings, at country resolution. In a validation study, test orders could confirm that these self-reported vendor locations are reliable as geographic indicators~\cite{rhumorbarbe2016buying}. As a consequence, the information architecture of darknet markets can allow researchers to broadly assess the trading geography of market offerings. It is commonly found that there is high spatial concentration: a small number of the same countries are responsible for the majority of global trades, regardless of the marketplace. Top trading countries include the US, the U.K., Australia, Germany, Holland, Canada, and China~\cite{christin2013traveling,barratt2014use,aldridge2016hidden,van2016sells,dolliver2016geographic,broseus2017geographical}. 

In an early exploratory spatial data analysis~\cite{dolliver2016geographic}, a study across four drug types attempts to relate darknet market listings on a single market to the known geography of the sites of production for these drugs. The geographic distributions are assessed visually and by identifying extreme outliers, and then manually compared with documented production geographies. Based on this comparison, the authors posit that the darknet trade of heroin and cocaine has a strong retail focus, in that most offerings were not sold from known producer countries. On the other hand, the markets for novel psychoactive substances and prescription drugs are more mixed, with a share of the offerings coming from producer regions~\cite{dolliver2016geographic}. 

A first study to assess the trading flows between vendor locations and potential destination countries finds some regional difference in trading geography, based on the listings of a single large market~\cite{broseus2017geographical}. Australian vendors primarily serve local customers, while vendors in other countries such as Holland, Germany, and China offer to trade internationally. It is further found that some countries are more specialised in their inventories than others. For example, offerings from Holland are comparatively more focused on ecstasy and stimulants, while those in other top trading countries are more diversified. In comparison to physical goods, listings for digital products are less likely to state a vendor location. 

\subsection{Knowledge Gaps} 
\label{sub:knowledge_gaps}
In principle, the presence of basic geographic indicators in darknet offerings allows to develop a more detailed understanding of the global production networks behind these trades. Yet to date, the literature offers no comprehensive empirical assessment of the spatial relationships between production, trade, and consumption of particular drug types on darknet markets. Research in crime science suggests that producers may also become participants on such markets, potentially removing previous trafficking structures. On the other hand, early studies suggest that a high spatial concentration of trades is largely shaped by the consumption geography of the traded goods, rather than the production geography. There presently is no analysis of how such relationships may vary across market platforms. How closely connected are production regions to these new online distribution channels? To what extent have old supply chains been fully replaced by darknet markets? 

\section{Research Questions} 
\label{sec:research_questions}
We seek to develop a more detailed understanding of the global production networks behind darknet market trades. In the present study, we focus on the spatial relationships between production, trade, and consumptions in the drug supply chain. 

\subsection*{Q1: Is the geography of darknet trading sited closer to the supply- or demand-sides of illegal drug production networks?} 
\label{sub:q1}
How well does the geography of darknet market vendor locations match the documented distribution of global production sites, or of global consumption? For certain drug types, comparing these geographies allows us to draw inferences about the underlying supply chain, from production, to trade, to consumption. Does increased local production of certain drugs lead to an increase in trade? Conversely, is trade demand-driven, and largely shaped by drug consumption in the local population?

Based on prior observations in the literature, we expect to find no significant link between production geography and trading volumes. Instead, we expect to find that market supply is largely demand-driven. This finding would suggest that darknet markets have not necessarily replaced prior supply chains, and are instead primarily novel forms of local distribution serving the `last mile'.

\subsection*{Q2: Do these relationships vary across markets? } 
\label{sub:q2}
It is feasible that there are `wholesale' marketplaces where producers predominantly serve distributors, and `retail' markets where distributors serve end users. For certain drug categories we expect to see this reflected in their trading geography, which may more closely reflect the geography of drug producers, or that of drug consumers. Can we identify such market specialisation, based on the geographic distribution of offerings? Or are the spatial relationships relatively constant across markets?

Although some markets may serve slightly different geographies, based on prior work it is likely that the global distribution pattern is largely comparable across markets. We expect to find a similar trading geography across most markets, situated closer to the consumption-side than the production-side.

\section{Methods} 
\label{sec:methods}
For the purpose of this study, we define darknet markets as two-sided markets that operate as Tor hidden service, connecting large numbers of vendors to large numbers of buyers, offering a wide range of licit and illicit products that are catalogued in a structured manner comparable to e-commerce platforms. We exclude single-vendor sites, and specialty markets for single product categories such as stolen credit cards, software exploits, and others.

We rely on two primary sources of evidence. First, we comprehensively survey the darknet market ecosystem in May 2017, and identify the largest platforms by inventory. Based on this survey, we select four large markets which represent approximately 80\% of the global inventory, and capture their full trading histories. This data represents a significant historic archive of illicit transactions, comprising approximately 1.5M trades. Second, using data by the UNODC we determine the geography of supply and demand for three plant-based drug types: cannabis, cocaine, and opiates. Using these two data sets, we produce a comprehensive statistical model of the economic geography of darknet market trades, incorporating trading activity across more than 50 countries. For each drug type, we build a regression model to explain the geography of its trade based on its production and consumption geography. 

\subsection{Scraping the Largest Markets} 
\label{sub:scraping_the_largest_markets}
We began with an initial survey of the ecosystem, using link lists published on the English-speaking internet in May 2017.\footnote[2]{\url{https://dnstats.net/}}\footnote[3]{\url{http://grams7enufi7jmdl.onion/markets}}\footnote[4]{\url{https://www.reddit.com/r/DNMSuperlist/wiki/superlist}}\footnote[5]{\url{https://www.deepdotweb.com/2013/10/28/updated-llist-of-hidden-marketplaces-tor-i2p/}} We identified 36 unique markets that fit our criteria. Of these, we further excluded platforms that had gone offline, forum-based platforms which do not provide structured catalogue listings, and platforms that required an invite. Of the remainder, 21 platforms provided inventory counters in their navigation structure, which allowed us to estimate the total size of their trading inventory. Special care was taken to include the most prominent sites that were featured across multiple link lists. 

Our survey revealed high centralisation: the largest market claimed to offer 60\% of the global inventory, and the largest 5 were responsible for almost 95\%. We selected four of the largest markets for inclusion in our study, covering 80\% of global listings: Alphabay, Hansa, Traderoute, and Valhalla. 
A fifth market (at 14\%) was not included, as its information architecture did not allow us to determine trading estimates on a per-product basis (see next paragraph).

For each of the four markets, we scraped a single snapshot of the full catalogue in late June to early July 2017. We pursued a variety of strategies to discover product pages: while the sequential identifiers used by three markets allowed us to discover hidden listings, a fourth market required a crawl of catalogue listings to gradually discover all listed products. To ensure completeness we observed platform downtimes, session expiry, and other intermittent failure modes, and retried requests when necessary. 

We estimated trading volumes based on buyer reviews for listed products, informed by prior work published in the literature~\cite{christin2013traveling,soska2015measuring}. We estimate one trade for every review posted on a market. Since buyer reviews are not mandatory on all markets, the resulting estimates serve as a lower bound of actual trades. This approach allowed us to establish significant trading histories: Alphabay and Traderoute displayed a full history of all reviews, Hansa displayed the last six months of reviews, and Valhalla the last three months. 

\begin{table}
  \centering
\begin{tabular}{ l | r r r }
\textbf{Market} & \textbf{Vendors} & \textbf{Items} & \textbf{Trades}\footnotemark[6] \\
\midrule
\textbf{Alphabay}   & 7,500 & 140,300 & 1,308,500 \\
\textbf{Hansa}      & 2,300 & 51,800  & 91,900 \\
\textbf{Traderoute} & 600   & 9,200   & 3,300 \\
\textbf{Valhalla}   & 1,100 & 30,400  & 9,100 \\
\midrule
\textbf{Total}      &       & 231,700 & 1,412,800 \\
\end{tabular}
\caption{Catalogue sizes of the four platforms included in the study, rounded to hundreds. Data from June/July 2017.}
\label{tab:market_summary}
\end{table}

\footnotetext[6]{Estimated from buyer reviews, refer to Section~\ref{sub:scraping_the_largest_markets} for details.}

In total, the collected data accounts for almost 1.5M trades. The resulting summary statistics of market catalogues and trading estimates are shown in Table~\ref{tab:market_summary}. Vendors provided a country of origin for approximately 80\% of the listings, which allowed us to establish a detailed trading geography. The top countries in number of trades are the USA (27\%), Great Britain (22\%), Germany (8\%), Australia (8\%), and Holland (7\%), together these `top five' account for more than 70\% of the global trade. The remaining 10\% are distributed across 114 countries. This distribution suggests a highly unequal trading geography: by comparison, in 2015 the same five countries only represented 7\% of the global population~\cite{un2017world}.

To prepare the data for analysis, we normalised the stated geographic location of vendors by mapping them to ISO 3166 country codes. We further identified all cannabis, cocaine and opiate products across markets based on product titles and market-specific product categories. Products categorised under terms such as `buds', `weed', `hash', `cannabis', `cannabis concentrates', or similar were labelled as $Cannabis$. Products whose categorisation or title contained `cocaine' were labelled $Cocaine$. Opiates were identified by mentions of `heroin', `morphine', or `opium', and were labelled $Opiates$. In all cases, we excluded synthetic variants and prescription drug preparations such as fentanyl, hydrocodone, and oxycodone, as well as listings of test kits for these drugs.

\subsection{Production and Consumption Geography} 
\label{sub:geography_of_production_and_consumption}
The annual UNODC World Drug Report collects detailed data about the global geography of production, trafficking, and consumption of several widely traded drug types. This includes country-level statistics that we can use as basis for our analysis.~\cite{unodc2017world} We extract two sets of data from this report. 

First, the consumption geography of different drugs is described by a report on drug use. This is a national estimate of the percentage of adults who have consumed a particular drug category in the preceding year, grouped into major drug categories. Methodologies for data collection vary by country and sample, but typically these are based on surveys among adult participants of age 15 and older. In cases where multiple survey results were available for a country and drug type, we selected the most recent result. Most of the information is sufficiently recent: 82\% of the selected records were from the last 10 years (2007-2016), and only 24\% from 2001-2005. Inspection of the data showed that these older reports contained countries from all major continents, and we do not expect that their inclusion introduces a systematic geographic bias.

Second, we derive information about the geography of drug production from a dataset of 490,000 significant drug seizures, available for the years 2010-2015. This data set records the presumed production country of each seized good, as stated in arrest reports. Unfortunately, seized drug volumes are reported in a variety of units that cannot be map to a single unit of measurement, including weights, volumes, tablet counts, and ampoules. Instead, we determine the number of seizures per country as a baseline estimate of local drug production activity. Some manual labelling was required to accommodate our analysis, as drug names in this data are not standardised and categorised to the same degree. Non-synthetic opiates that may have been preprocessed, but are not yet in medicine form were labelled as $Opiates$. Cocaine in raw and processed form but not yet in medicine form was labelled $Cocaine$. All organic forms of cannabis were labelled $Cannabis$, excluding synthetic variants. For every drug type, we computed the overall number of seizures that occurred, aggregated by the presumed country of production.

For both data sets we normalised country designations, mapping them to ISO 3166 country codes when there was a single and unambiguous mapping. Data for the U.K. presented a special case, as it was reported separately for England and Wales, Scotland, and Northern Ireland. For population prevalence data, we computed a weighted average for UK based on recent population sizes~\cite{ONS2017pop}. For seizures, we aggregated the national counts.

The data comes with some limitations. These relatively comprehensive data sets depend on national reports, which are not available for all countries. Information about drug production is derived from arrest records and seizures, which are not collected for all countries. The U.K. and Canada only make information about seizures available for some drug types, and for Australia this data is not available at all. While Mexico is reportedly one of the world's largest cocaine producers, it does not feature in the global cocaine seizures data. Additionally, the seizures data is likely skewed by differences in local law enforcement practices. Despite these limitations, this is likely the most detailed source of information available on the global drug economy~\cite{unodc2017worldmeth}.

\subsection{Feature Vectors} 
\label{sub:feature_vectors}
From these data sets we compute a feature vector per drug type, each containing country-specific measurements for three variables:
\begin{itemize*}
  \item $Seizures$: the number of significant seizures of a drug where the country is the presumed site of production.\footnote[7]{While local enforcement practices can vary widely, this measure can serve as a baseline estimate of local drug production. Refer to Section~\ref{sub:geography_of_production_and_consumption} for details.}
  \item $Trades$: the number of trades on darknet markets by vendor location, as estimated by the number of buyer reviews.
  \item $UseInPop$: the share of the population that has consumed the drug in the last year.
\end{itemize*}

\begin{figure}
  \centering
  \includegraphics[width=\linewidth]{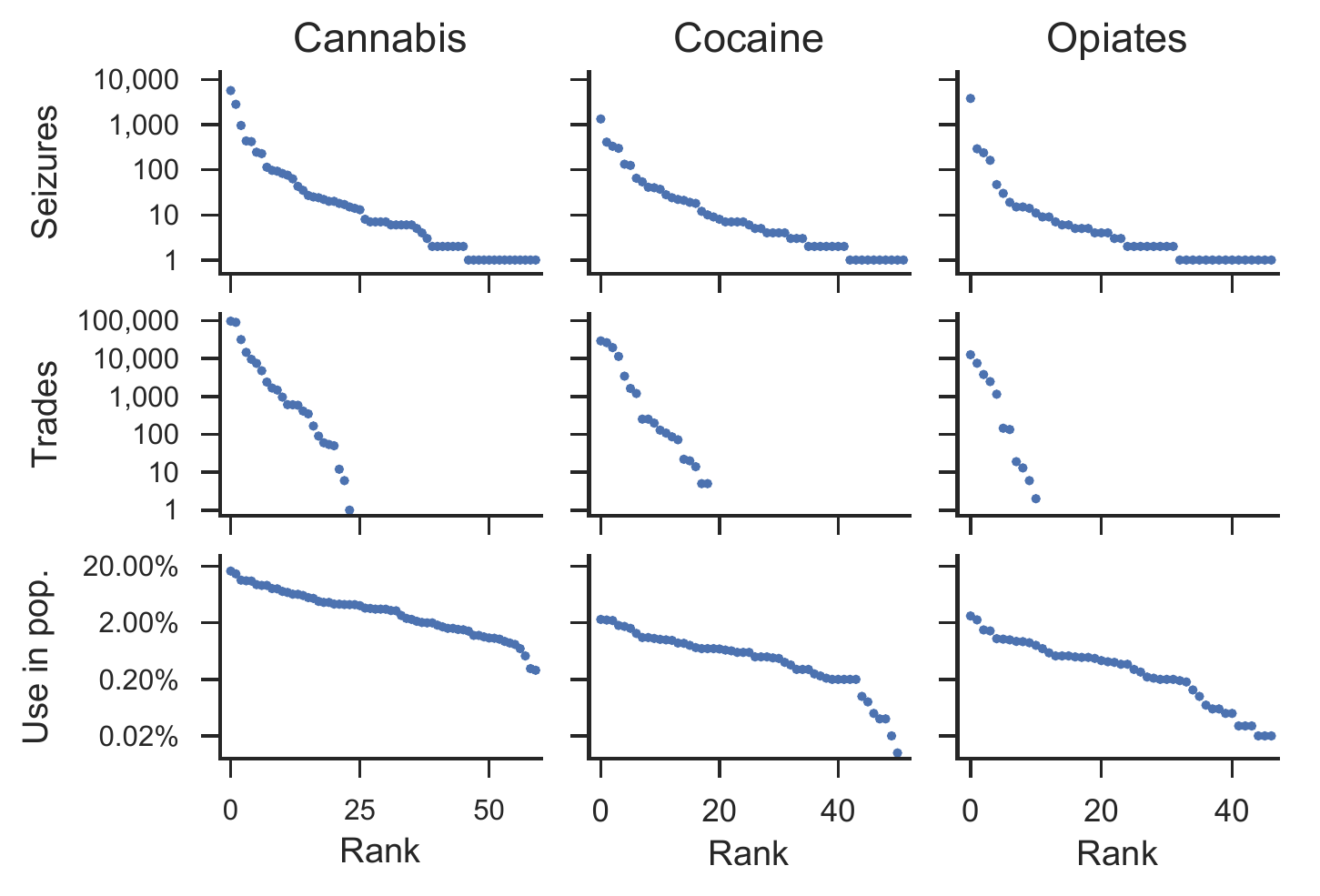}
  \caption{Feature distribution per drug type. Visualised as ranked country measurements, using a logarithmic scale.}
  \label{fig:feature_dist}
\end{figure}

For each feature vector, we only included countries for which all three variables can be determined. We regard darknet market data as a comprehensive survey of the global ecosystem, and countries that did not feature in trades are treated as having 0 sales. On the other hand, data about national production and consumption are regarded as samples, and countries that are not included in UNODC reports are treated as unknown measurements, and cannot be included in our analysis. Such exclusions are acceptable as long as a sufficiently large number of major trading countries is represented in the data. After exclusion, the feature vector for cannabis includes complete measurements for 60 countries, for cocaine 52 countries (both include US, GB, DE, and NL but exclude AU), and for opiates 47 (with US, DE, and NL, but excluding GB and AU.) The data is available for download\footnote[8]{\url{https://dx.doi.org/10.5287/bodleian:E9jknXek6}}. 

\begin{figure}
  \centering
  \includegraphics[width=\linewidth]{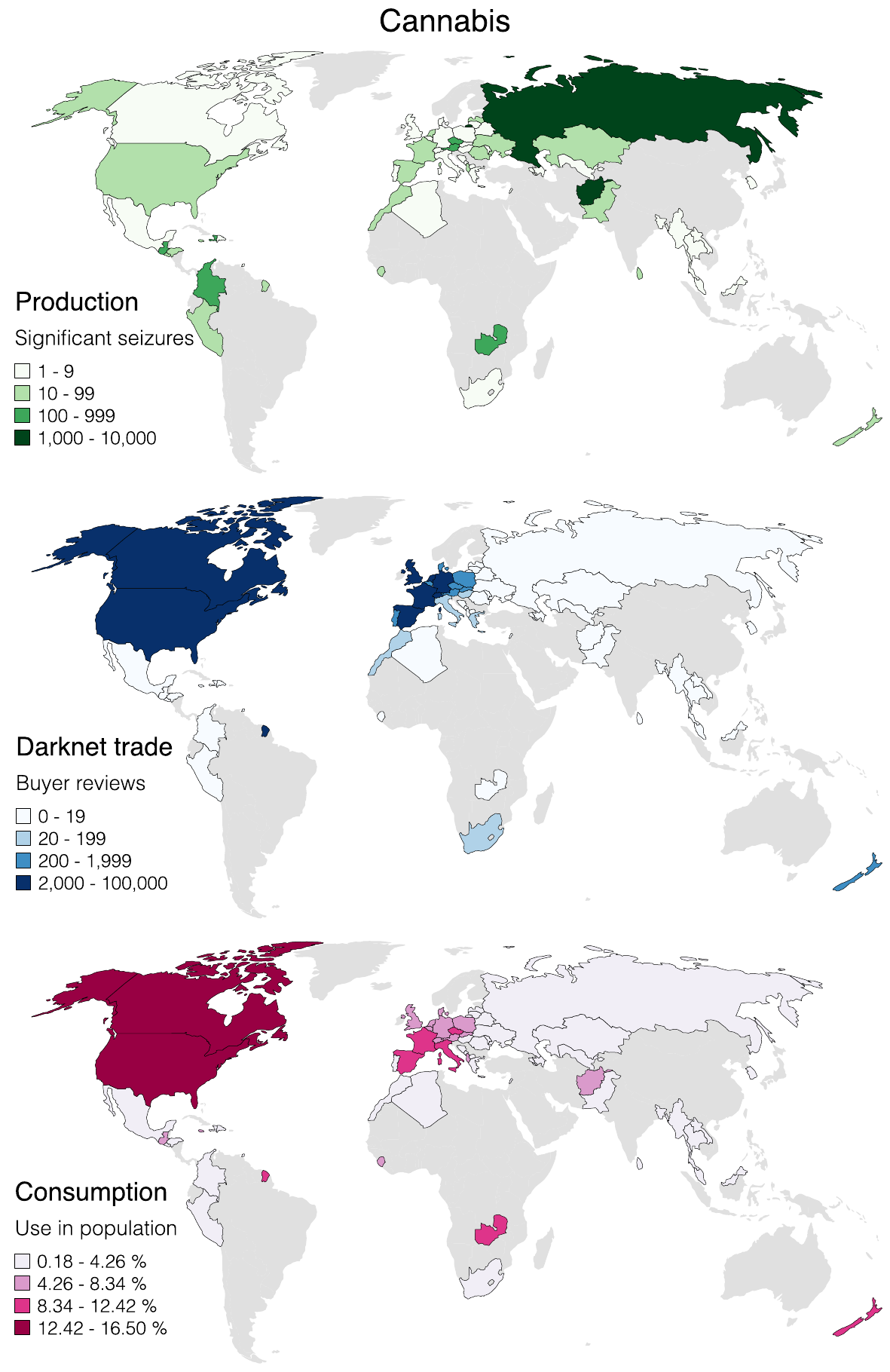}
  \caption[Global distribution of cannabis production, trade, and consumption.]{Global distribution of cannabis production, trade, and consumption\footnotemark[9].}
  \label{fig:map_cannabis}
\end{figure}

\footnotetext[9]{Production is measured as the number of seizures of significant drug shipments, and attributed to the country of production. Trade is measured as the number of drug sales on darknet markets, as estimated from buyer reviews. Consumption is measured as the percentage of a population who report drug use in the last year.}

\begin{figure}
  \centering
  \includegraphics[width=\linewidth]{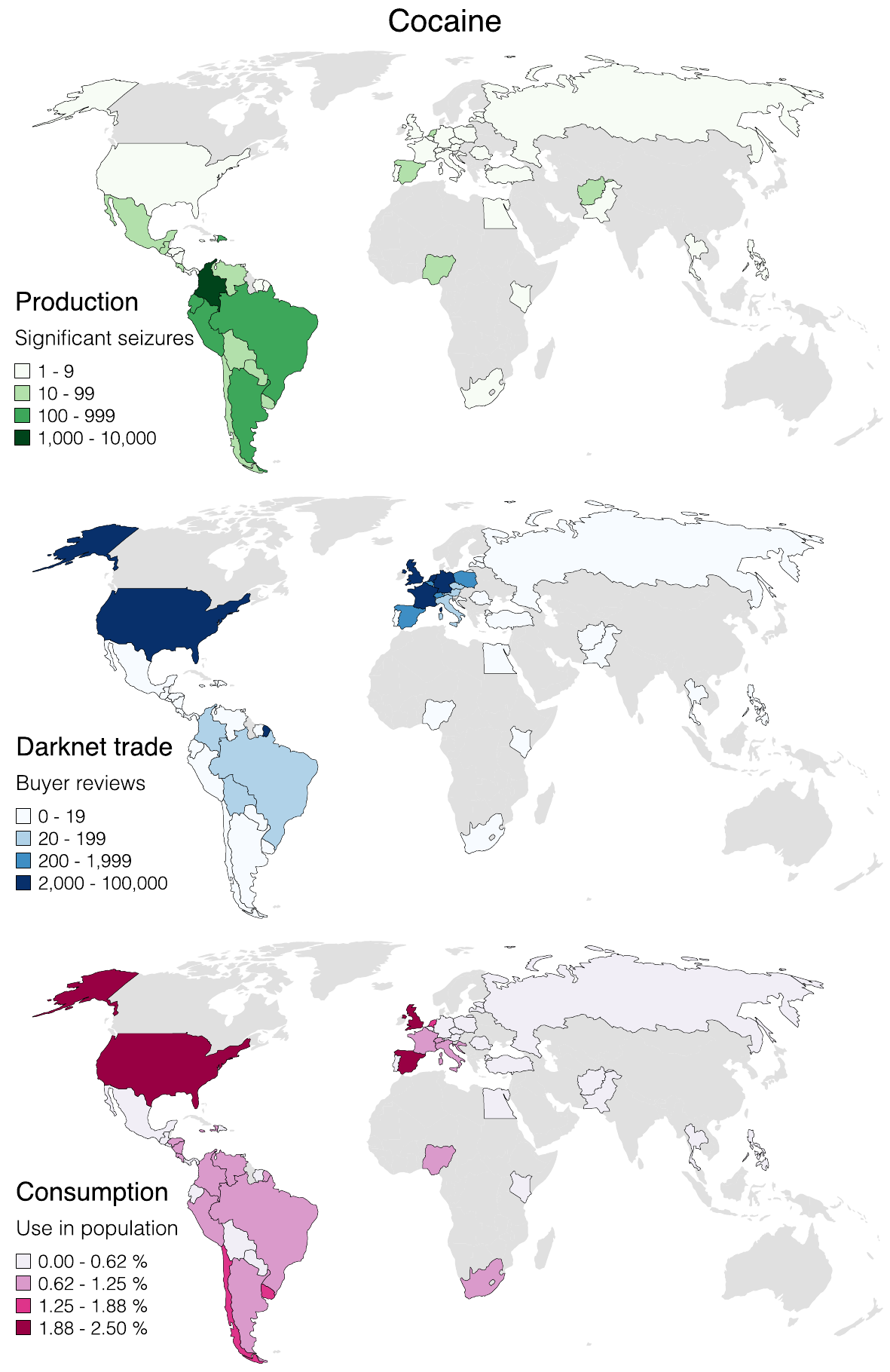}
  \caption[Global distribution of cocaine production, trade, and consumption.]{Global distribution of cocaine production, trade, and consumption.}
  \label{fig:map_cocaine}
\end{figure}

\begin{figure}
  \centering
  \includegraphics[width=\linewidth]{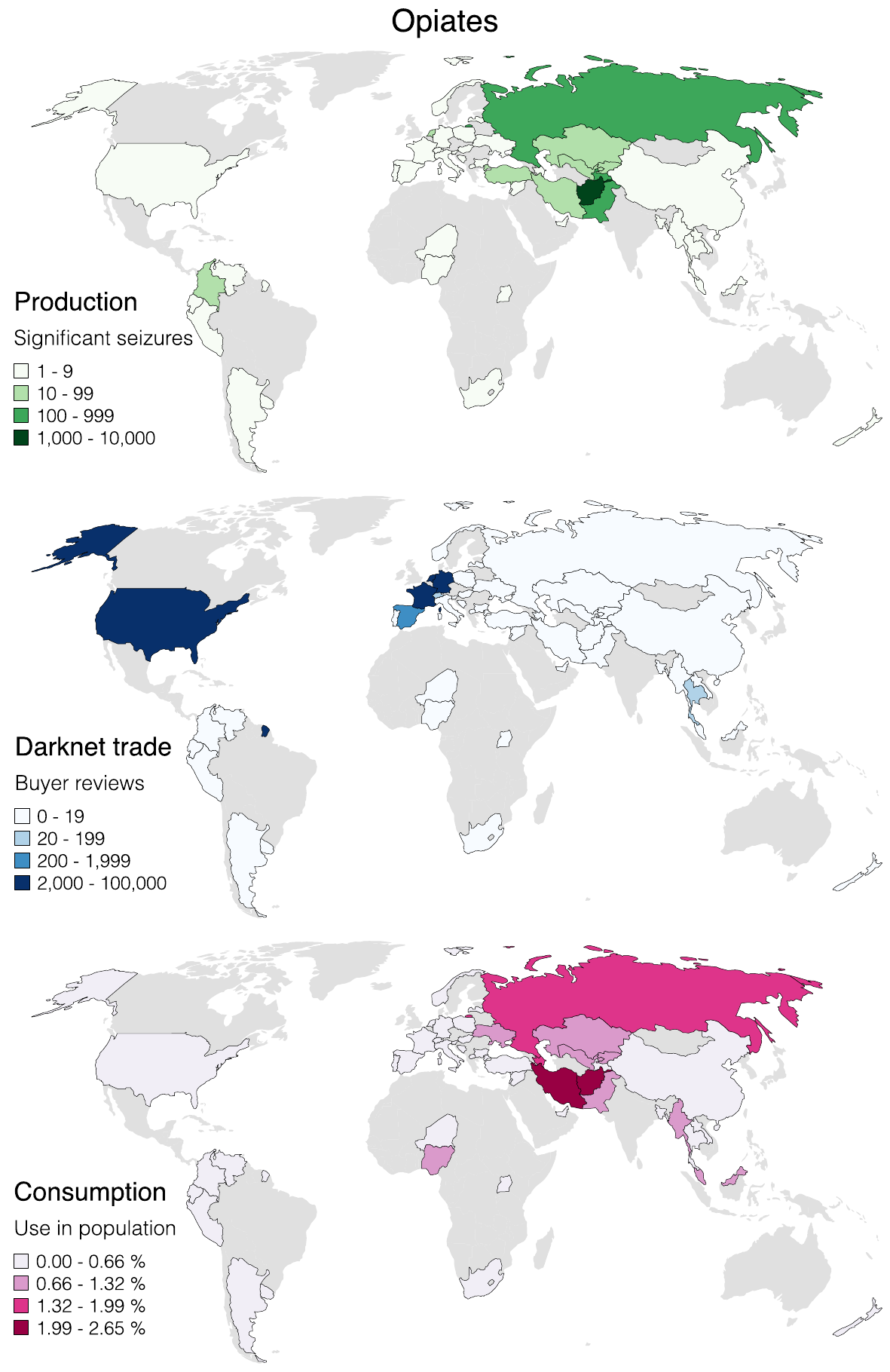}
  \caption[Global distribution of opiate production, trade, and consumption.]{Global distribution of opiate production, trade, and consumption.}
  \label{fig:map_opiates}
\end{figure}

The variable distribution across the three feature vectors is shown in Figure~\ref{fig:feature_dist}. As none of the measures is normally distributed, we derived standardised and log-transformed variants for all feature vectors. In our analysis we will primarily report model outcomes for the normalised measures, but we always also compute outcomes for the standardised and log-transformed feature vectors to confirm the result, and we will state when these differ. 

According to these measures, consumption and production geography are correlated for cocaine (Spearman correlation coefficient $\rho_S=0.29$, $p<0.05$). This is supported visually by the maps in Figure~\ref{fig:map_cocaine}, which show that several South American countries are both producers and consumers of this drug. For cannabis and opiates, there is no significant relationship between the geographies of production and consumption, suggesting they are largely independent. This is confirmed by the maps in Figure~\ref{fig:map_cannabis} and Figure~\ref{fig:map_opiates}. We observe low condition indices for all feature vectors (cannabis: 1.2, cocaine: 1.3, opiates: 1.9), which shows low multicollinearity across the features. This also suggests that the correlated inputs for the cocaine feature vector are unlikely to bias regression outcomes.

In addition to these global feature vectors, we produced four market-specific variants that cover the same countries. We normalised the variables in each feature vector to address their differences in measurement ranges. This allows us to compare the proportional relationships of a consumption-side and production-side geography between drug types and markets, regardless of their respective trading volumes.

\subsection{Analysis} 
\label{sub:analysis}
To address Q1, we first visually related feature variable distributions by means of a parallel coordinates plot. We then compared the data statistically by means of a pairwise correlation analysis, then with an OLS linear regression model to explain the distribution of $Trades$ using $Seizures$ and $UseInPop$. For the market comparison in Q2, we further compute individual regression models per market. All outcomes of these analyses are reported in the following sections.

\begin{figure}
  \centering
  \includegraphics[width=0.95\linewidth]{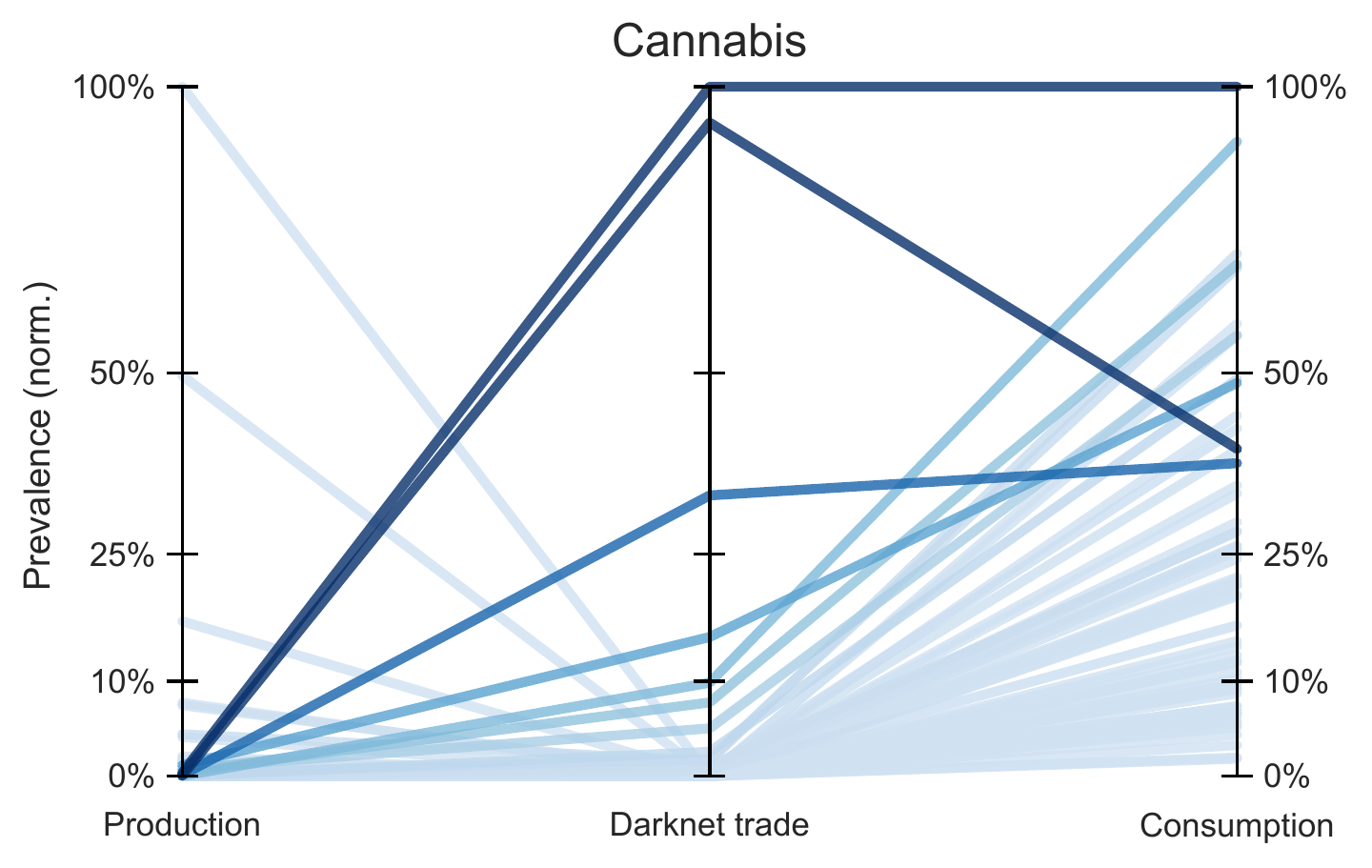}
  \caption{Parallel coordinates plot: national prevalence of cannabis, across normalised measures of production, trade, and consumption, on a logarithmic scale. Darker shades indicate countries with higher trading volumes. The plot illustrates that for the top trading countries, national consumption of the drug is high, while national production is low.}
  \label{fig:pcor_cannabis}
\end{figure}

\begin{figure}
  \centering
  \includegraphics[width=0.95\linewidth]{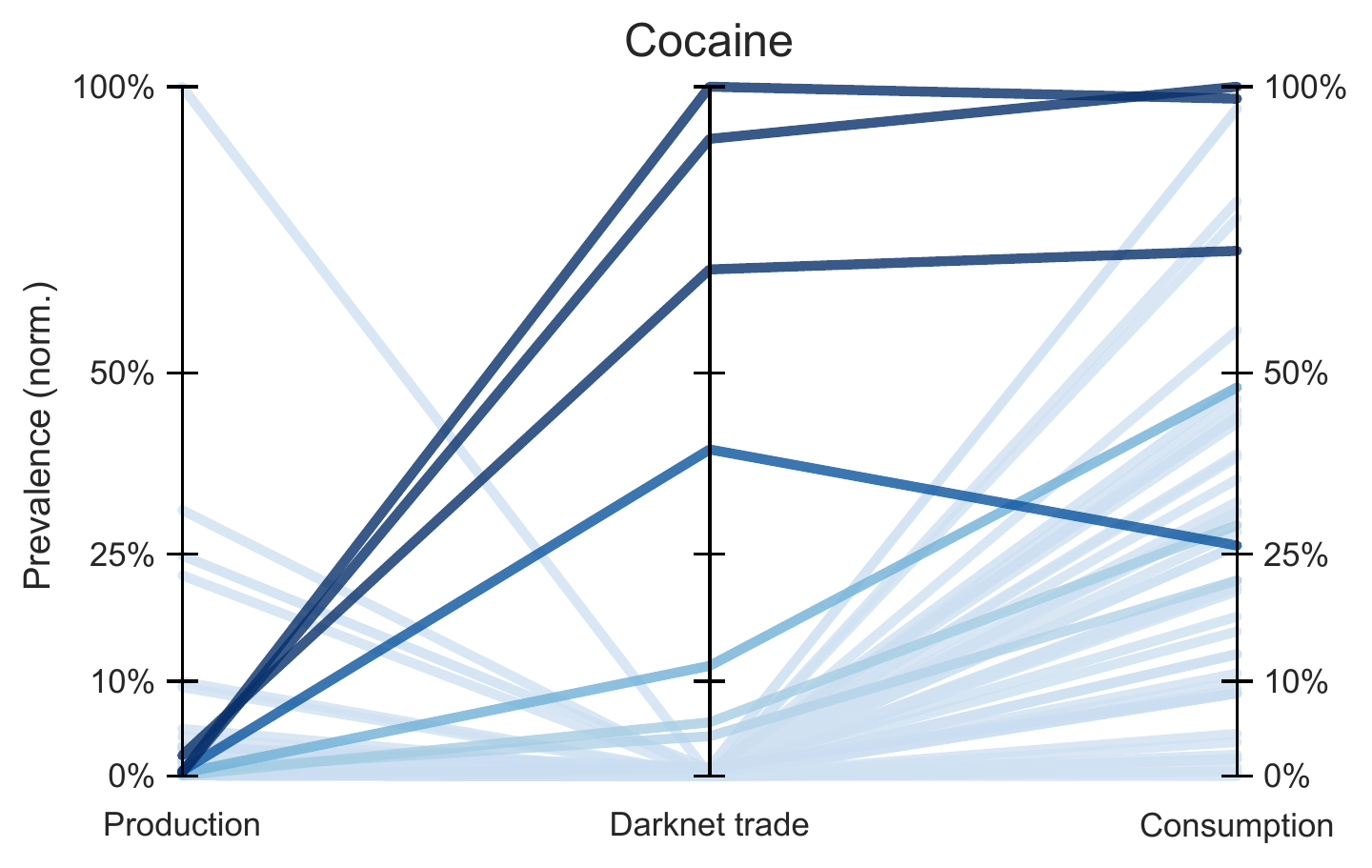}
  \caption{National prevalence of cocaine. For the top trading countries, there is a clear relationship between trade and national consumption, but not national production.}
  \label{fig:pcor_cocaine}
\end{figure}

\begin{figure}
  \centering
  \includegraphics[width=0.95\linewidth]{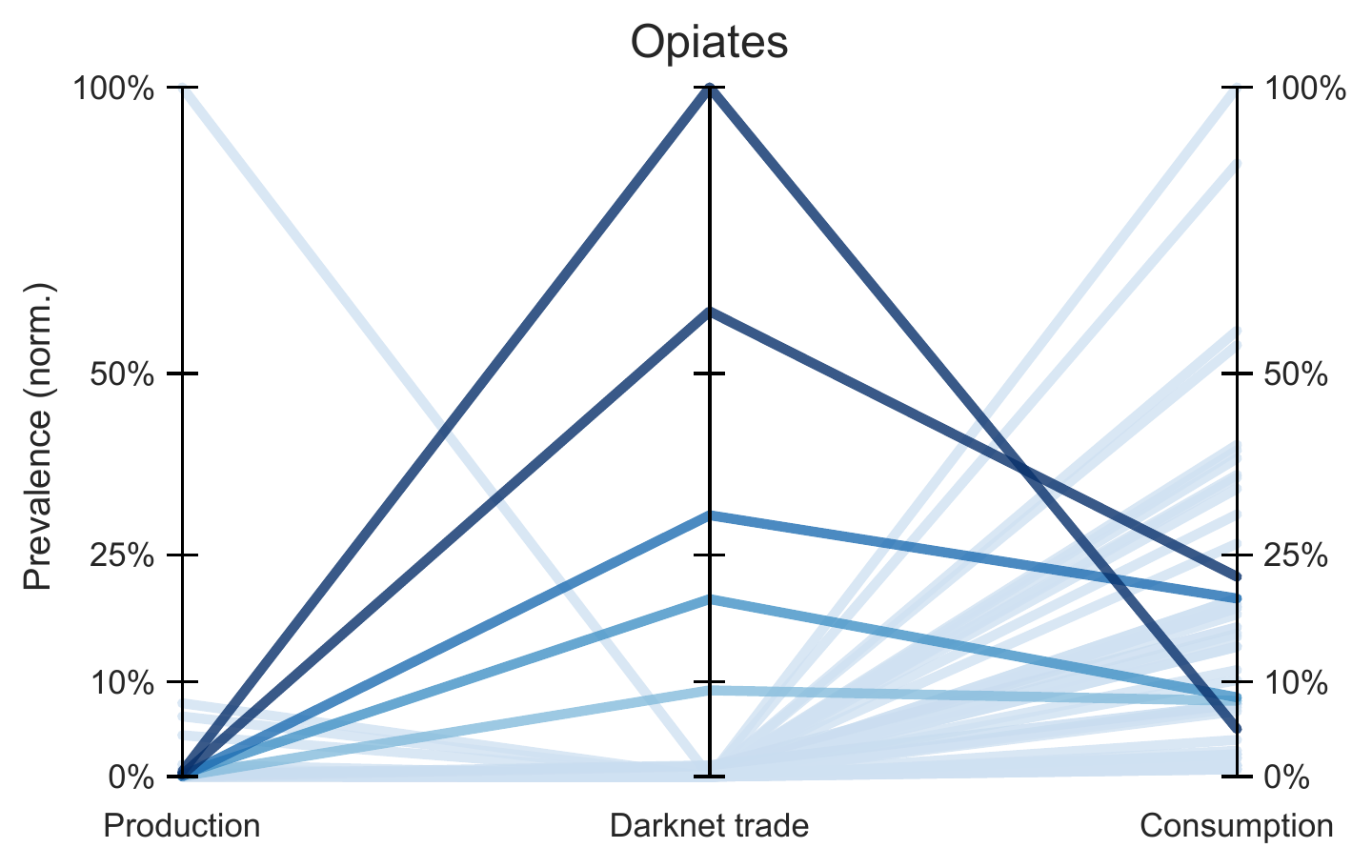}
  \caption{National prevalence of opiates. For this drug type the relationships are less clear: darknet trade is largely independent of both national production and consumption.}
  \label{fig:pcor_opiates}
\end{figure}

\section{Results} 
\label{sec:findings}
\subsection{Q1: Overall Trading Geography} 
\label{sub:q1_a}
The feature vectors for each drug type are visualised as parallel coordinate plots in Figures~\ref{fig:pcor_cannabis}-\ref{fig:pcor_opiates}. In these plots, the three feature variables are represented using three vertical axes, ranging from 0-100\% on a logarithmic axis, and placed alongside each other for visual comparison. Each country is represented as a line drawn across the three axes, so that the line's anchor point on each axis corresponds to the country's measurement for the respective variable.

The plot for cannabis in Figure~\ref{fig:pcor_cannabis} suggests a relationship between local darknet trade and drug consumption, but not local production of the drug. It further illustrates that this relationship holds not just for the top trading countries, but also for countries with lower trading volumes. The plot for cocaine in Figure~\ref{fig:pcor_cocaine} suggests a similar spatial relationship. 
On the other hand, the spatial relationships for opiates are less clear: as Figure~\ref{fig:pcor_opiates} shows, there is no strong relationship between local production and local trade, or local consumption and trade. The map in Figure~\ref{fig:map_opiates} illustrates this relationship further.
While peak opiate consumption primarily happens in the Middle East, Russia, and Asia, darknet trading volumes of opiates are highest in the same countries as the other two drugs.

These pairwise relationships are confirmed with a correlation analysis. For cannabis, there is significant correlation between national trade and consumption ($\rho_S=0.66$, $p<0.001$), but not trade and production. For cocaine, national trade and consumption are correlated ($\rho_S=0.36$, $p<0.01$), but not trade and production. For opiates, neither variable pairs are correlated.
An OLS linear model confirms these relationships. Table~\ref{tab:q1_models} reports the regression coefficients for the cannabis model ($adjusted\ R^2=0.19$) and for cocaine ($adjusted\ R^2=0.33$). Both report a significant relationship between local trade and consumption, but not production. A model for opiates was not significant for any of the covariates. These relationships hold for the standardised and log-transformed variants of the three feature vectors.

\begin{table}
  \centering
\begin{tabular}{ l l | r | r | r }
  \textbf{Aspect} & \textbf{Variable} & \textbf{Cannabis} & \textbf{Cocaine} & \textbf{Opiates} \\
\midrule
  Production  & \textbf{$Seizures$}  & (-0.08)        & (-0.11)       & (0.02) \\
  Consumption & \textbf{$UseInPop$}  & \textbf{0.38}  & \textbf{0.48} & (-0.08) \\
  & \textbf{$Intercept$} &  (-0.05)       & (-0.07)       & (0.06) \\
\end{tabular}
\caption{Regression coefficients for three drug models to explain national trading volumes ($Trades$) based on national production ($Seizures$) and consumption ($UseInPop$) of each drug. Coefficients in bold are significant with $p < 0.001$, those in brackets are not significant.}
\label{tab:q1_models}
\end{table}

\subsection{Q2: Market-Specific Trading Geography} 
\label{sub:q2_a}
We first assessed the trading geography for every market by means of a pairwise correlation analysis between drug trade and production, and trade and consumption. The results are shown in Table~\ref{tab:q2_corr}, they each reflect the same spatial relationships as encountered for the overall trade. Results for the corresponding market-specific regression models are reported in Table~\ref{tab:q2_models}, all models are highly consistent with the global model. In other words, for cocaine and cannabis, all four markets appear to be located in demand-side geographies, but not supply-side geographies. As before, the comparison for opiates was inconclusive. 

\begin{table}
  \centering
\begin{tabular}{ l l | r | r | r }
  \textbf{Market} & \textbf{Variable} & \textbf{Cannabis} & \textbf{Cocaine} & \textbf{Opiates} \\
\midrule
  Alphabay  & \textbf{$Seizures$} & (-0.08)        & (0.11) & (-0.09) \\
            & \textbf{$UseInPop$} & \textbf{0.67}  & 0.35   & (-0.11) \\
\midrule
  Hansa     & \textbf{$Seizures$} & (-0.04)        & (0.08) & (-0.04) \\
            & \textbf{$UseInPop$} & \textbf{0.56}  & 0.41   & (-0.03) \\
\midrule
  Traderoute& \textbf{$Seizures$} & (0.02)         & (0.06) & (0.20) \\
            & \textbf{$UseInPop$} & \textbf{0.47}  & 0.37   & (-0.04) \\
\midrule
  Valhalla  & \textbf{$Seizures$} & (0.01)         & (0.03) & (0.01) \\
            & \textbf{$UseInPop$} & \textbf{0.56}  & 0.34   & (-0.06) \\
\end{tabular}
\caption{Pairwise Spearman correlation coefficients between $Trades$ and $Seizures$ or $UseInPop$ across the four markets, demonstrating that the spatial relationships are broadly consistent across markets. Coefficients in bold are significant with $p < 0.001$, those in brackets are not significant, all others are significant with $p < 0.02$.
}
\label{tab:q2_corr}
\end{table}

\begin{table}
  \centering
\begin{tabular}{ l l | r | r | r }
  \textbf{Market} & \textbf{Variable} & \textbf{Cannabis} & \textbf{Cocaine} & \textbf{Opiates} \\
\midrule
  Alphabay  & \textbf{$Seizures$}  & (-0.08)        & (-0.11)       & (0.02) \\
            & \textbf{$UseInPop$}  & \textbf{0.38}  & \textbf{0.48} & (-0.08) \\
            & \textbf{$Intercept$} &  (-0.05)       & (-0.07)       & (0.06) \\
\midrule
  Hansa     & \textbf{$Seizures$}  & (-0.05)        & (-0.12)       & (-0.01) \\
            & \textbf{$UseInPop$}  & \textbf{0.31}  & \textbf{0.43} & (-0.05) \\
            & \textbf{$Intercept$} &  (-0.06)       & (-0.06)       & (0.06) \\
\midrule
  Traderoute& \textbf{$Seizures$}  & (-0.11)        & (-0.07)       & (-0.04) \\
            & \textbf{$UseInPop$}  & \textbf{0.44}  & \textbf{0.32} & (0.01) \\
            & \textbf{$Intercept$} &  (-0.04)       & (-0.05)       & (0.03) \\
\midrule
  Valhalla  & \textbf{$Seizures$}  & (-0.08)        & (-0.08)       & (0.03) \\
            & \textbf{$UseInPop$}  & \textbf{0.34}  & 0.27          & (-0.08) \\
            & \textbf{$Intercept$} &  (-0.03)       & (-0.03)       & (0.05) \\
\end{tabular}
\caption{Individual regression models for the four markets. Coefficients in bold are significant with $p < 0.001$, those in brackets are not significant, all others are significant with $p < 0.02$.
}
\label{tab:q2_models}
\end{table}

\section{Discussion} 
\label{sec:discussion}
Our analysis is based on a systematically collected and large corpus, containing a full trading history of the four largest markets in mid-2017, comprising 1.5M estimated trades. According to an initial survey, this represents 80\% of the global darknet market inventory.
We assessed spatial patterns visually, and confirmed our observations using rigorous statistical methods. This choice of methods allows us to interpret the contextual nature of our findings in some detail. 
We will first summarise and contextualise the key findings, and then discuss their implications.

First, we confirm that \emph{trading is strongly focused on a small number of highly-active countries}. 70\% of global trades are attributable to the `top five' of USA, U.K., Australia, Germany, and Holland. (A further 20\% are not associated with any country of origin.) In comparison, the same countries only represent 7\% of the global population. This strong spatial concentration has previously been observed by others~\cite{christin2013traveling,barratt2014use,aldridge2016hidden,van2016sells,dolliver2016geographic,broseus2017geographical}, and is illustrated here with some further clarity in the form of detailed maps and diagrams. 

Second, we find a clear and simple relationship: \emph{trade of cocaine and cannabis on darknet markets is highest in peak consumer countries for these drugs}, rather than peak producer countries. While it remains possible that some producers of such drugs also sell directly on darknet markets, the vast share of darknet offerings are provided by redistributors who are in closer proximity to end users. 

Third, \emph{this relationship cannot be confirmed for opiates}. For this drug type, trade is distributed differently than both production and consumption. The maps in Figure~\ref{fig:map_opiates} and parallel coordinates plot in Figure~\ref{fig:pcor_opiates} illustrate this: while opiate consumption is highest in the Middle East, Russia, and Asia, opiate trading volumes are highest in the `top five' countries. This suggests that more complex models are needed to explain the trading geography of these drugs. We make some suggestions for future work in Section~\ref{sub:future_work}.

Fourth, \emph{neither of the drug types is shipped from peak producer locations} to a significant degree, including opiates. This further emphasises that market activity is firmly taking place in consumer regions: the trading is local or regional, rather than global.

We posit that these findings have high external validity: our models for cocaine and cannabis are consistent for four major markets across more than 50 countries, and our observations of the geographic distribution of trades are consistent with previously published findings for other marketplaces. 

\subsection{Implications} 
\label{sub:implications}
The spatial pattern observed here confirms that for the small number of highly active countries, darknet market trading is focused on the `last mile': vendors are situated closer to places of consumption than places of production. This makes it unlikely that darknet markets are removing or replacing prior supply chains. Rather, they appear to reconfigure the last mile of the supply chain for certain consumer countries, leaving existing trafficking routes from production sites largely untouched. Darknet vendors likely still rely on some of the same producers and traffickers involved in the conventional drug trade. In other words, darknet markets are not disintermediators with global reach, but rather provide a novel form of retail distribution for certain consumer countries, potentially offering new forms of risk reduction for buyers and sellers. 

Our findings further suggest that platform adoption is largely shaped by extrinsic factors, rather than platform-specific aspects. The four markets show similar spatial arrangements, which suggests they inhabit similar roles in the supply chain: neither of these markets is situated more closely to producers, instead all are firmly situated on the demand side. Overall, this also suggests that their adoption is largely driven by existing external demand, rather than new demand brought about by these new markets. This finding is not surprising, given that darknet markets only represent a small fraction of the global drug trade. 
As a consequence, it is unlikely that current darknet markets significantly alter the global drug production network. Similarly, market closures by law enforcement are unlikely to effect changes to the public consumption of illicit drugs, instead new markets will likely emerge to fulfil the existing large demand. Recent research has already documented such an ecological robustness among darknet markets~\cite{duxbury2017building,decary2017police,ladegaard2017we}.

Our findings have further implications for the use of darknet market data as indicator measures. Researchers who seek to use trading data to model societal behaviours should do so carefully. We can observe that in some specific cases, market data may be a meaningful proxy for national levels of drug consumption. However we also find that this varies widely by drug type and geography. For use in models, it is advisable to limit the use of darknet data to well-specified scenarios, and then make efforts to carefully ground these measurements. We should further emphasise that the data is systematically biased in a number of aspects, including the particular demographic profile of participants~\cite{van2013surfing,barratt2014use,barratt2016safer,barratt2016if,van2016characterising}.

\section{Conclusion} 
\label{sec:conclusion}
Are darknet markets slowly reorganising the global drug trade? In a large-scale empirical study, we determined the darknet trading geography of three plant-based drugs across the largest markets, and compare it to the global footprint of production and consumption for these drugs. This is the first study to analyse the economic geography of darknet markets in such a comprehensive empirical manner, and our findings allow us to draw new inferences about the nature of the underlying supply chain. We cannot find evidence for a significant producer presence across any of the drug types or marketplaces. Instead, we present strong evidence that cannabis and cocaine vendors are primarily located in a small number of highly active consumer countries, suggesting that darknet markets primarily inhabit the role of local retailers serving the `last mile' for certain regions. This likely leaves old trafficking routes intact. By comparison, a model to explain trading volumes of opiates is inconclusive. Our findings further suggest that the geography of darknet market trades is primarily driven by existing consumer demand, rather than new demand fostered by these markets. We propose that this presents a theoretical contribution that helps characterise the economic geography of darknet markets with new specificity. Our study provides new evidence that helps situate these platforms within the global production networks of the illicit drug trade.

\subsection{Future Work} 
\label{sub:future_work}
The proposed model is too simple to explain the trading geography of opiates, and possibly other drug categories. Further work is needed to develop models that better characterise the economic geography of these products.
Similarly, there are unresolved questions about the drivers behind darknet market adoption. What can explain the geographic focus on a small number of consumer countries? Additional factors need to be considered that can mediate such adoption, including internet use, local drug policy and enforcement practices, and other socio-economic indicators. Finally, models of the darknet market global production network should consider the retail nature of these markets in more detail. There are suggestions in the literature that these platforms may not be exclusively focused on selling to end users, but may also be bulk suppliers to buyers who then become offline distributors~\cite{aldridge2014not,aldridge2016hidden,baravalle2016mining}. Are darknet markets best considered as retail centres, or as brokers for regional wholesale suppliers? 

\begin{acks}
This work was supported by \grantsponsor{ati}{The Alan Turing Institute}{https://www.turing.ac.uk} under the \grantsponsor{epsrc}{EPSRC}{https://www.epsrc.ac.uk} grant \grantnum{epsrc}{EP/N51012}.
Joss Wright is partially funded as a Turing Fellow under Turing Award Number \grantnum{ati}{TU/B/000044}, and Mark Graham is partially funded as a Turing Fellow under Turing Award Number \grantnum{ati}{TU/B/000042}.
We further acknowledge the support of Google for funding the ``Economic Geographies of the Darknet'' project at the Oxford Internet Institute.
\end{acks}

\bibliographystyle{ACM-Reference-Format}
\balance
\bibliography{web2018dittus} 

\end{document}